\documentclass{nature}

\usepackage[latin1,utf8]{inputenc}
\usepackage{graphicx,color}
\usepackage{amsmath,amssymb,amsfonts}
\usepackage{stackrel}
\usepackage[normalem]{ulem}
\usepackage{tabularx}
\newcolumntype{Y}{>{\centering\arraybackslash}X}
\usepackage{booktabs}
\AtBeginDocument{
    \heavyrulewidth=.08em
        \lightrulewidth=.05em
        \cmidrulewidth=.03em
        \belowrulesep=.65ex
        \belowbottomsep=0pt
        \aboverulesep=.4ex
        \abovetopsep=0pt
        \cmidrulesep=\doublerulesep
        \cmidrulekern=.5em
        \defaultaddspace=.5em
}
\usepackage{xcolor}
\usepackage{enumitem}
\usepackage{hyperref}
\hypersetup{
    colorlinks=true,
        linkcolor={blue!80!black},
        citecolor={blue!80!black},
        urlcolor={blue!80!black} 
}

\providecommand{\openone}{\leavevmode\hbox{\large1\kern-
7.3pt\normalsize1}}
\newcommand{\be}{\begin{equation}}
\newcommand{\ee}{\end{equation}}
\newcommand{\ba}{\begin{eqnarray}}
\newcommand{\ea}{\end{eqnarray}}

\newcommand{\fig}{Fig.~}

\newcommand{\lk}{\left[}
\newcommand{\rk}{\right]}

\newcommand{\dint}[3]{\int_{#2}^{#3}\!\mathrm{d}#1 \,}

\newcommand{\DeltaLogEps}{\Delta \!\ln \epsilon}
\newcommand{\ratioP}{p/p_\text{FD}}

\begin{document}

\title{Evidence for quark-matter cores in massive neutron stars}

\author{Eemeli Annala$^1$, Tyler Gorda$^2$, Aleksi Kurkela$^{3,4}$, Joonas N\"attil\"a$^{5,6,7}$ \& Aleksi Vuorinen$^1$}

\maketitle

\begin{affiliations}
    \item Department of Physics and Helsinki Institute of Physics, P.O.~Box 64, FI-00014 University of Helsinki, Finland
    \item Department of Physics, University of Virginia, Charlottesville, Virginia 22904-4714, USA
    \item Theoretical Physics Department, CERN, Geneva, Switzerland 
    \item Faculty of Science and Technology, University of Stavanger, 4036 Stavanger, Norway
    \item Physics Department and Columbia Astrophysics Laboratory, Columbia University, 538 West 120th Street, New York, 10027, USA
    \item Center for Computational Astrophysics, Flatiron Institute, 162 Fifth Avenue, New York,  10010, USA
    \item Nordita, KTH Royal Institute of Technology and Stockholm University, SE-10691 Stockholm, Sweden
\end{affiliations}

\begin{abstract}

\boldmath

The theory governing the strong nuclear force, Quantum Chromodynamics, predicts that at sufficiently high energy densities hadronic nuclear matter undergoes a deconfinement transition to a new phase of quarks and gluons\cite{Shuryak:1980tp}. Although this has been observed in ultrarelativistic heavy-ion collisions\cite{Gyulassy:2004zy,Andronic:2017pug}, it is currently an open question whether quark matter exists inside neutron stars\cite{Lattimer:2004pg}. By combining astrophysical observations and theoretical \emph{ab-initio} calculations in a model-independent way, we find that the inferred properties of matter in the cores of neutron stars with mass corresponding to 1.4 solar masses ($M_\odot$) are compatible with nuclear model calculations. However, the matter in the interior of maximally massive, stable neutron stars exhibits characteristics of the deconfined phase, which we interpret as evidence for the presence of quark-matter cores. For the heaviest reliably observed neutron stars\cite{demorest:2010bx,antoniadis:2013pzd} with mass $M \approx 2M_\odot$, the presence of quark matter is found to be linked to the
behaviour of the speed of sound $c_s$ in strongly interacting matter. If the conformal bound $c^2_s \leq 1/3$\cite{Cherman:2009tw} is not strongly violated, massive neutron stars are predicted to have sizable quark-matter cores. This finding has important implications for the phenomenology of neutron stars, and affects the dynamics of neutron star mergers with at least one sufficiently massive participant.

\unboldmath

\end{abstract}

Observations of neutron stars (NSs) inform us about the properties of matter inside their cores in an indirect way. To translate them to statements about NS matter requires modeling strongly interacting matter all the way from the crust to the highest densities reached inside the stars. The lack of accurate first-principles predictions at densities beyond the nuclear matter saturation (baryon number) density $n_0 \approx 0.16/{\rm fm}^3$ has so far prevented the determination of the phase of matter inside NS cores, and it is unlikely that the question will be answered based on gravitational wave (GW) data alone, at least in the near future\cite{Bauswein:2018bma}. Nevertheless, recent observations are beginning to offer empirical constraints so strong that a  model-independent approach to the problem has become feasible. 

The equation of state (EoS) of NS matter---the relation $p(\epsilon)$ between the pressure and energy density of beta-equilibrated matter interacting under quantum chromodynamics (QCD) at temperature $T = 0$---is known in two opposing limits. From the well-studied NS crust region\cite{Fortin:2016hny} to the density $n_\text{CET} \equiv 1.1 n_0$, where matter resides in the hadronic-matter phase, modern nuclear-theory machinery, such as chiral effective field theory (CET), provides the EoS to good precision, currently better than $\pm 24\%$\cite{Gandolfi:2009fj,Tews:2012fj}. In the opposite limit of very high densities, perturbative-QCD (pQCD) techniques, rooted in high-energy particle phenomenology and built on deconfined quark and gluon degrees of freedom\cite{Kurkela:2009gj,Gorda:2018gpy}, become accurate, providing the quark-matter EoS to the same accuracy at densities $n \gtrsim 40 n_0 \equiv n_\text{pQCD}$.  

In the above two limits, QCD matter is known to exhibit markedly different properties. High-density quark matter is approximately scale invariant, or conformal, whereas in hadronic matter the number of degrees of freedom is much smaller and, additionally, scale invariance is violated by the breaking of chiral symmetry. These qualitative differences are reflected in the values taken by different physical quantities. The \emph{speed of sound} takes the constant value $c_s^2 = 1/3$ in exactly conformal matter, and slowly approaches this number from below in high-density quark matter\cite{Kurkela:2009gj}. By contrast, in hadronic matter the quantity varies considerably: below saturation density, CET calculations indicate $c_s^2\ll 1/3$, while at higher densities most hadronic models predict max$(c_s^2)\gtrsim 0.5$. The \emph{polytropic index} $\gamma \equiv \mathrm{d}(\ln p)/\mathrm{d}(\ln \epsilon)$ on the other hand obtains the value $\gamma = 1$ in conformal matter, while both CET calculations and hadronic models generically predict $\gamma \approx 2.5$ around and above saturation density. Finally, the number of degrees of freedom is reflected in the \emph{pressure normalized by that of free quark matter} (the Fermi-Dirac, or FD, limit), $\ratioP$\cite{Kurkela:2009gj}. This quantity obtains ${\mathcal O}(0.1)$ values in CET calculations and hadronic models, while pQCD predictions typically fall inside the range $[0.5,0.8]$ around $n=n_\text{pQCD}$.

In the intermediate-density range $n_\text{CET} < n < n_\text{pQCD}$, where NS cores lie, a robust model-independent approach is to introduce a set of basis functions to interpolate the EoS, thus creating an ensemble of all viable NS-matter EoSs\cite{Hebeler:2013nza,Kurkela:2014vha,Annala:2017llu,Most:2018hfd}. To remove possible bias originating from the choice of basis functions, we have used multiple interpolation methods, finding consistent EoS and NS mass-radius (MR) regions, as reviewed in the Methods section (and shown in Extended Data \fig{1}). Owing to this agreement, in the following we present results from a new ``speed-of-sound'' interpolation method we have here developed, which has the added benefit of keeping track of the stiffness of the EoS and allows for arbitrarily strong crossover transitions, tantamount to discontinuous first-order transitions. We impose the following two robust astrophysical constraints on the EoS: the requirement of supporting a $1.97M_\odot$ NS\cite{demorest:2010bx,antoniadis:2013pzd} and that the tidal deformability $\Lambda$ for a 1.4$M_\odot$ star obeys $70 < \Lambda(1.4 M_\odot ) < 580$\cite{TheLIGOScientific:2017qsa,Abbott:2018exr}. In total, we have analyzed approximately 570,000 EoSs, displayed in  \fig{1}, to which we apply a mild smoothness condition in some individual analyses, as discussed in the Methods section (and shown in Extended Data \fig{2}).

%%%%%%%%%%%%%%%%%%%%%%%%%%%%%%%%%%%%%%
\begin{figure}
\caption{
\label{finalfig1}
\textbf{Range of allowed neutron-star-matter equations of state.} The bands have been generated by superimposing large numbers of individual EoSs generated with the speed-of-sound interpolation method introduced in this paper. The color coding refers to the maximal value that the speed of sound squared $c_{s}^{2}$ reaches at any density. For comparison, the black lines stand for the different hadronic EoSs we have obtained from refs.~\cite{Lattimer:2000nx,Gandolfi:2011xu,Fortin:2016hny}. Finally, the light blue regions correspond to the CET and pQCD EoSs of \cite{Hebeler:2013nza,Kurkela:2009gj}, and the rough location of the deconfinement transition in hot quark-gluon plasma, $\epsilon_\text{QGP}$, is indicated for illustrative purposes.}{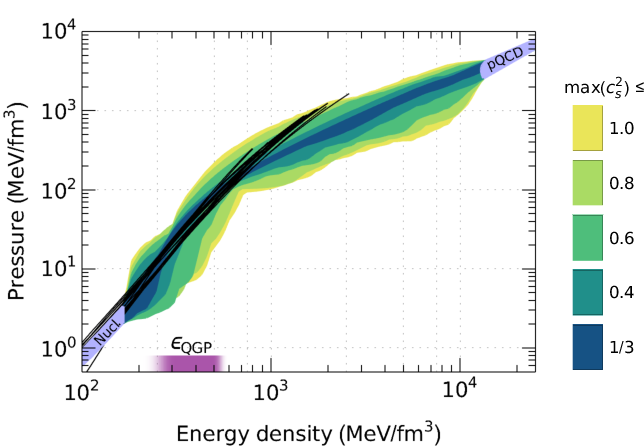}{0.7\textwidth}
\end{figure}
%%%%%%%%%%%%%%%%%%%%%%%%%%%%%%%%%%%%%%

Having the ensemble of interpolated EoSs at our disposal, we can determine the allowed behaviors of different physical quantities. In \fig{2}, panel a, we display a 3d rendering, where a representative sample of our EoSs is represented by thin black lines, all starting from a region characterized by the CET EoS and ending in a pQCD one. For comparison, we also include a large set of nuclear matter EoSs (thick black lines, corresponding to the hadronic EoSs of \fig{1}, obtained from refs.~\cite{Lattimer:2000nx,Gandolfi:2011xu,Fortin:2016hny}, of which we have discarded those incompatible with our observational constraints. The interpolated EoSs follow a non-trivial trajectory: at low densities, they follow a trend set by the nuclear EoSs but later deviate from it, signalling a change in the underlying physics. This transition corresponds to the change of (the logarithmic) slope of $p(\epsilon)$ visible around $\epsilon_c \approx 400-700$ MeV/fm$^3$ in \fig{1}, which roughly coincides with the energy density inside free nucleons and to the location of the deconfinement transition at high temperatures\cite{Borsanyi:2013bia,Bazavov:2014pvz}. At even higher densities, the interpolated EoSs approach the pQCD predictions for $c_s^2$, $\ratioP$, and $\gamma$\cite{Kurkela:2009gj}.

%%%%%%%%%%%%%%%%%%%%%%%%%%%%%%%%%%%%%%%%%%%%%%%%%%%%%%
\begin{figure}
\caption{
\label{finalfig2}
\textbf{Characterization and microscopic interpretation of the equations of state.}
\textbf{a,} A 3d rendering of trajectories of a representative subset of interpolated (thin black lines) and hadronic (thick black lines) EoSs in a space spanned by the polytropic index $\gamma$, the pressure ratio $\ratioP$, and the squared speed of sound $c_s^2$. The solid blue and empty cyan diamonds mark the centers of $1.4M_\odot$ NSs, and the solid red and empty magenta circles denote the same for $M_\text{max}$ stars. 
The thick light blue line finally denotes the region of the 3d space spanned by the high-density pQCD EoS, and the region occupied by low-density matter is indicated by the label ``Nucl''. \textbf{b,c,} Projection of the 3d image to the $\ratioP$--$\gamma$ (\textbf{b}) and $c_s^2$--$\gamma$ (\textbf{c}) planes. In panel \textbf{c}, the light-blue star indicates the high-density conformal pQCD limit.}{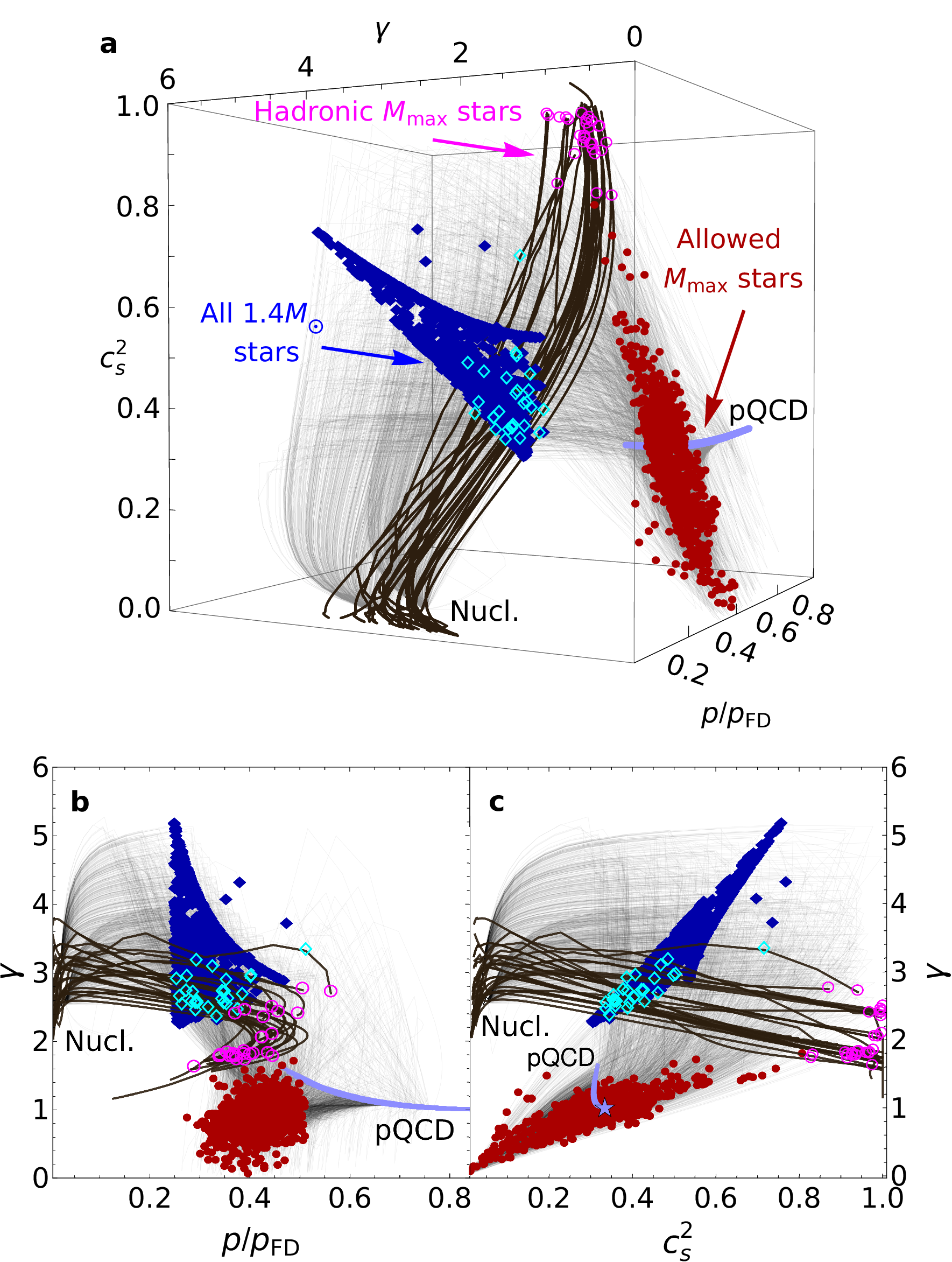}{0.95\textwidth}
\end{figure}
%%%%%%%%%%%%%%%%%%%%%%%%%%%%%%%%%%%%%%%%%%%%%%%%%%%%%%

On each of the EoS lines, the density reached in the centers of $1.4M_\odot$ NSs is marked with solid blue (interpolated EoSs) or empty cyan (nuclear EoSs) diamonds. The significant overlap between the two distributions shows that the material properties inside $1.4M_\odot$ stars are consistent with a description in terms of hadronic degrees of freedom. The same is, however, not true for the majority of maximally massive stars (filled red and empty magenta circles), for which the two families of points are clearly separated in \fig{2}.
(Panels b and c of \fig{2} 
display two 2d projections of the 3d plot.) This discrepancy shows that the material properties in the centers of the largest NSs are not consistent with the presence of hadronic matter except for a small number of EoSs with very large values of $c_s^2$. 
As demonstrated in Extended Data \fig{3}, the vast majority of these points lie to the right of the kink on the EoS band, confirming the likely existence of quark cores inside the $M_\text{max}$ stars.

In order to make the above observations somewhat more quantitative, it would clearly be valuable to establish a connection between the physical phase of QCD matter and its EoS. To this end, we note that our results from \fig{2} 
and the distinct values the polytropic index obtains in nuclear and quark matter calculations both suggest using the values of $\gamma$ as a good approximate criterion. Given that $\gamma=1.75$ is both the average between its pQCD and CET limits and very close to the minimal value the quantity obtains in viable hadronic models (see \fig{2} 
and our discussion in the Methods section), we are led to choose the following criterion for separating hadronic from quark matter: given an interpolated EoS, the smallest density from which $\gamma<1.75$ continuously to asymptotic densities is identified with the onset of quark matter. We emphasize, however, that this is only an approximate rule to guide our analysis, and not a robust or rigorous result.

As a first application of the above criterion, for NSs with $M=1.4M_{\odot}$ we find that the central polytropic index always satisfies $\gamma\gtrsim 2$, implying that the stars are composed of hadronic matter as expected. In contrast, maximally massive stars have $\gamma$ values much closer to unity, indicating that they typically contain quark matter. In \fig{3}, 
we display the sizes of quark cores in the latter NSs. The core has a significant extent, $M_{\rm core} > 0.25 M_\odot$, for all those EoSs that satisfy $c_s^2 < 0.5$. However, for extreme EoSs in which the speed of sound almost reaches that of light, the core may be significantly smaller or even absent.  Note that this is consistent with the fuller picture given by \fig{2} 
mentioned above.

If the maximal value of $c_s^2$ exceeds $0.7$, we find a small class of EoSs where even maximally massive stars do not contain quark cores according to our criterion. We find that each of these EoSs exhibits {an interval in $\epsilon$ where $\gamma < 0.5$, which destabilizes the star. This corresponds to a rapid change in the EoS, and is practically indistinguishable from a first-order phase transition.} The minimal latent heat { (\emph{i.e.}~the extent of the interval with $\gamma < 0.5$)} required for the destabilization is $(\Delta \epsilon)_{\rm lat} > 130$ MeV/fm$^3$, corresponding to a relative discontinuity $(\Delta \epsilon)_{\rm lat}/ \epsilon > 0.2$ at the beginning of the transition. We thus find that in order for {all stable NSs} to be composed of hadronic matter alone, the EoS must both significantly violate the conformal limit and feature a sufficiently strong phase transition.  Finally, two-solar-mass stars contain a quark core for all EoSs that satisfy $c_s^2 < 0.4$, irrespective of the properties of the phase transition; for subconformal  EoSs{, featuring $c_s^2<1/3$ at all densities,} the radius of the core $R \approx 6.5$~km is roughly half of the entire star's radius. By contrast, if the EoS supports substantially higher maximal masses $M_{\max} > 2.25 M_\odot$, quark cores are absent in $2M_\odot$ stars  (see Extended Data \fig{4}).

%%%%%%%%%%%%%%%%%%%%%%%%%%%%%%%%%%%%%
\begin{figure}

\caption{
\label{finalfig3}
\textbf{The size of the quark core.} Predictions for the radii and masses of the quark cores in maximally massive NSs are displayed. The maximal value that the speed of sound squared $c^2_s$ reaches in each individual EoS is indicated by color-coding the corresponding points. Points corresponding to lower $c_s^2$ values are drawn on top of those corresponding to higher ones. The NS in the inset visualizes a $12$-km, $2M_\odot$ star with a $6.5$-km quark core, built with a subconformal ($c_s^2<1/3$) EoS.}{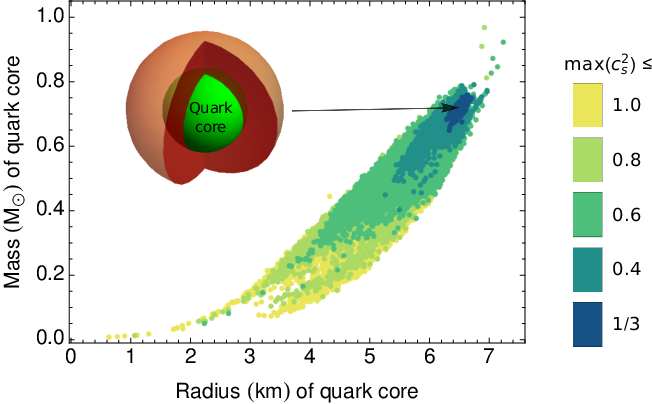}{0.7\textwidth}
\end{figure}
%%%%%%%%%%%%%%%%%%%%%%%%%%%%%%%%%%%%%

In conclusion, our model-independent analysis {has demonstrated} that the existence of quark cores in massive NSs should be considered the standard scenario, not an exotic alternative: for all stars to be made up of hadronic matter, the EoS of dense QCD matter must be truly extreme. This view is also consistent with  recent NS radius measurements, which are compatible with the larger radii predicted by the less extreme EoSs (see discussion under Methods, and Extended Data \fig{5}).
Note, however, that our analysis does not preclude the possibility of massive, purely hadronic stars even with less extreme EoSs, as quark cores may appear only at very high masses, even beyond $2M_\odot$. Nuclear matter EoSs that predict purely hadronic $2M_\odot$ stars (see, \emph{e.g.},\cite{Muther:1987xaa,Akmal:1998cf,Typel:2009sy,Fattoyev:2010mx,Steiner:2012rk}) may therefore be compatible with our results until very high densities.

The existence of quark cores in at least some NSs{, and that} the nucleation of quark matter begins so close to the maximum-mass limit, may have very interesting observable consequences. In NS mergers, 
the core may lead to shock waves reflecting from the quark-hadron interface inside hypermassive NSs. This may be particularly amplified if the conformal limit is strongly violated in hadronic matter, leading to large differences in the speeds of sound between the phases. In addition, the onset of the transition may give rise to dissipation during the merger in the form of a large effective bulk viscosity that may lead to an enhanced damping of the ringdown. Importantly, both of these have the potential to lead to observable effects in GW signals from NS mergers and the associated EM counterparts. 

Finally, our results  are systematically improvable with more  observations.
For example, there are several candidates for NSs with very large masses (see \emph{e.g.}~ref.~\cite{Cromartie:2019kug}).
If even one of these stars turns out to have a mass significantly larger than $2M_\odot$, this would impose strong new constraints on the EoS and \emph{e.g.}~imply that the conformal bound must be 
broken. Similarly, with many binary-NS merger observations currently recorded by LIGO/Virgo, the current limits on the tidal deformability will inevitably become tighter, enabling additional improvements to our analysis. With these advances and the road map laid out in our work, further significant progress in understanding the nature of ultra-dense matter inside NSs can be expected in the near future.

\begin{addendum}
    \item We would like to thank Ingo Tews for helpful discussions and collaboration in early stages of this work. In addition, we acknowledge useful discussions with Paul Chesler, Niko Jokela, Simonetta Liuti, Anton Rebhan, Sanjay Reddy, Urs Wiedemann, and Kent Yagi. The work of E.A., T.G., and A.V. has been supported by the European Research Council, grant no.~725369 and by the Academy of Finland, grant no.~1322507. In addition, E.A. gratefully acknowledges support from the Finnish Cultural Foundation and T.G. from the U.S.~Department of Energy Grant No.~DE-SC0007984.
\end{addendum}

\begin{methods}

%%%%%%%%%%%%%%%%%%%%%%%%%%%%%%%%%%%%%%%%%%%
\subsection{New speed-of-sound interpolation}
\label{AppCs}
%%%%%%%%%%%%%%%%%%%%%%%%%%%%%%%%%%%%%%%%%%%

The starting point of the new interpolation method is to consider the squared speed of sound $c_{s}^{2}$ as a function of the baryon chemical potential $\mu$, and use this quantity to construct all other thermodynamic functions, in particular the pressure $p(\mu)$. In practice, the speed of sound is first integrated from the CET matching point $n_\text{CET}=1.1n_0$ to higher densities to give the baryon density
\begin{equation}
n(\mu) = n_\text{CET} \exp \left[ \int_{\mu_\text{CET}}^{\mu}\! \frac{\mathrm{d}\mu'}{\mu' c_s^2(\mu')} \right], \label{eq:n_from_cs2}
\end{equation}
where $\mu_\text{CET}$ is the baryon chemical potential corresponding to the density $n_\text{CET}$, that is, $n_\text{CET} \equiv n(\mu_\text{CET})$. This result is then further integrated to arrive at the pressure,
\begin{equation}
p(\mu) = p_\text{CET} + n_\text{CET} \dint{\mu'}{\mu_\text{CET}}{\mu} \exp \left[ \int_{\mu_\text{CET}}^{\mu'}\! \frac{\mathrm{d}\mu''}{\mu'' c_s^2(\mu'')} \right], \label{eq:p_from_cs2}
\end{equation}
where $p_\text{CET} \equiv p(\mu_\text{CET})$.

The above relations must be solved numerically in general, but in the following simple case that we have implemented in our analysis, they may be dealt with analytically. Namely, we first take sequence of $N_{p}$ pairs
\begin{equation}
\{(\mu_i , c_{s,i}^2)\}_{i = 1}^{N_{p}}, 
\end{equation}
with $\mu_{1} = \mu_{\text{CET}}$, $\mu_{N_{p}} = 2.6$~GeV, and $\mu_{i-1} < \mu_{i} < \mu_{i+1}$ for all other $i$. We then construct a $c_s^2$ curve as a piecewise-linear function connecting these points; that is, for each $i = 1,\ldots,N_{p} - 1$, and for $\mu \in [\mu_{i} , \mu_{i + 1}]$,
\begin{equation}
c_s^2(\mu) = \frac{(\mu_{i+1} - \mu) c_{s,i}^{2} + (\mu - \mu_{i}) {c_{s,i+1}^{2}}}{\mu_{i+1} - \mu_{i}}. 
\end{equation}
At the matching points $\mu_{1}$ and $\mu_{N_{p}}$, we require $p$ and $c_{s}^{2}$ to match the corresponding values given by the CET and pQCD EoSs, respectively. In addition, we take $n$ to be continuous at each matching point, but note that our construction allows for EoSs that mimic discontinuous first order transitions arbitrarily closely. 

For a given $N_{p}$, we have $N_{p} - 2$ independent matching chemical potentials $\mu_{i}$ and $N_{p} - 2$ independent speed-of-sound points ${c_{s}^{2}}_{i}$, from which one of both is determined through matching to the high-density EoS, leaving $2N_{p} - 6$ parameters for given low- and high-density EoSs. If we instead write this in terms of the number of interpolating segments $N \equiv N_{p} - 1$, the result becomes $2 N - 4$. This is one free parameter fewer than the number of free parameters needed to define a polytropic EoS composed of the same number of segments\cite{Annala:2017llu}.

The above procedure is used to construct individual EoSs by choosing $N = 3,4,5$, and then randomly picking values for the matching points $\mu_i$, speeds of sound $c_i$, and the pQCD parameter $X_\text{pQCD}$\cite{Fraga:2013qra}. 
The parameter values are taken from uniform distributions $\mu_{i} \in [\mu_{\text{CET}}, 2.6 \, \mathrm{GeV}]$, ${{c_{s}^{2}}_{i} \in (0,1)}$, $X_\text{pQCD} \in [1,4]$, in addition to which we choose roughly the same number of the ``hard'' or ``soft'' nuclear EoSs of ref.~\cite{Hebeler:2013nza}. Finally, we vary the extreme EoSs in the $\epsilon$, $p$ plane within each $c_s^2$ band plotted in our paper, to ensure that we satisfactorily probe the size of these regions. This leads to the ensemble studied above, which consists of approximately 570,000 individual EoSs. Roughly 160,000 of these fulfill the astrophysical constraints described in the main text, while approx.~70,000 of the allowed EoSs contain at least one first order phase transition. We have carefully made sure that these ensemble sizes are sufficiently high, so that our results are stable with respect to increasing the number of EoSs.

{Finally, we note that while the interpolation method described above is genuinely new, a number of related articles have recently appeared, in which the NS matter EoS has been constructed starting from the speed of sound\cite{Tews:2018chv,Tews:2018kmu,Greif:2018njt,Baym:2019iky,Bai:2019jtl,Bai:2019jqo}. While most of these works introduce a nontrivial ansatz function for the quantity, thus being more restrictive than our approach, in ref.~\cite{Tews:2018chv}\ the speed of sound is allowed to behave in a more general way. The main difference between the EoS bands constructed in this reference and our current work originates from our high-density pQCD constraint, which effectively forces the EoS to be softer at high densities.}

\subsection{Comparison of different interpolations}
\label{AppComparison}

To quantify the potential bias introduced into our results by the selection of the speed-of-sound interpolation method, we compare our EoS and MR ensembles to ones obtained with the following two schemes:
\begin{enumerate}
\item A piecewise polytropic interpolation of the pressure as a function of baryon density, \\ ${p_{i}(n)=\kappa_i n^{\Gamma_i}}$.
\item A spectral interpolation of the {adiabatic} index $\Gamma(p) = \frac{\epsilon(p) + p}{p} \lk \frac{\mathrm{d} \epsilon}{\mathrm{d}p} \rk^{-1}$ in terms of Chebyshev polynomials.
\end{enumerate}
Both of these interpolation methods have been abundantly discussed in the literature\cite{Hebeler:2013nza,Kurkela:2014vha,Annala:2017llu,Most:2018hfd,Lindblom:2010bb,Lindblom:2018rfr,Abbott:2018exr}.  

We construct the EoS bands corresponding to each of the three interpolation methods, implementing the astrophysical constraints listed in the main text. To make the EoS families comparable to each other, we have not only made sure that the ensembles are of roughly similar size, but have in addition chosen the numbers of free parameters in the EoSs approximately equal. For the piecewise-polytropic interpolation, we allow up to 4 independent segments\cite{Annala:2017llu} (amounting to 5 free parameters), while for the spectral interpolation proposed by Lindblom\cite{Lindblom:2010bb}, we use Chebyshev polynomials of degree 5 (4 free parameters). Finally, for the speed-of-sound interpolation, we use up to 5 independent segments (6 free parameters) in this comparison. In each case, we have randomly generated large ensembles of interpolation functions, ensured that the resulting EoSs are causal and thermodynamically consistent, and in the end discarded those EoSs that are in disagreement with the observational constraints introduced in the main text. Again, we add no explicit first order transitions to the EoSs, but allow continuous transitions that are arbitrarily strong, thus closely mimicking discontinuous phase transitions.

Our conclusion from the comparing the constructed EoSs (see Extended Data \fig{1}, 
panel a) is that the speed-of-sound and polytropic interpolations produce nearly identical results, while the spectral interpolation leads to a somewhat more constrained band. 
This fact is not surprising, considering that the spectral method does not build on piecewise-defined interpolating functions, so that the resulting EoSs are smooth by construction and unable to describe very sharp and rapid changes in the EoSs.

The families of MR curves obtained by integrating the Tolman-Oppenheimer-Volkoff (TOV) equations using the above three ensembles of EoSs also largely indicate agreement between the methods (see Extended Data \fig{1}, 
panel b). The minimal and maximal radii for a fixed mass agree well between the different interpolations, with the spectral interpolation occupying a slightly more restricted area for low-mass stars with $M<1.2 M_\odot$.  
The agreement between different interpolations also persists as a function of tidal deformability: Constraining $\Lambda(1.4M_\odot)$ according to $70<\Lambda(1.4M_\odot)< 580$\cite{Abbott:2018exr}, we find that the different interpolations still give similar maximal radii as functions of the NS mass as long as  $M\gtrsim 1.4 M_\odot$. In particular, the maximal radii at $M=1.4 M_\odot$ are in excellent quantitative agreement between the different interpolation methods, as was to be expected from the previously observed tight correlation between NS radii and tidal deformabilities\cite{Annala:2017llu}. Considering stars with smaller masses, we observe that the speed-of-sound and piecewise-polytropic interpolations allow EoSs that are extremely hard at low densities, leading to large radii $R\approx 14$~km for $M\approx M_\odot$, but rapidly soften at larger densities, such that for $M=1.4M_\odot$ the radii are smaller and consistent with the upper limits for the tidal deformability. Again, since the spectral method leads to smoother interpolations, it is natural that it does not allow these rapidly changing EoSs.

Another difference between the interpolation schemes is that the polytropic interpolation does not allow for as massive stars as the other two. We attribute this to the fact that in order to achieve very large maximal masses, the EoS needs to stay very stiff, $c_s \approx 1$, throughout an extensive density window, which is difficult to realize with polytropic interpolation functions.  This difference between different interpolations is somewhat ameliorated when  upper limits are placed on the tidal deformability.

%%%%%%%%%%%%%%%%%%%%%%%%%%%%%%%%%%%%%
\subsection{Polytropic index and its relation to the phase of QCD matter}
\label{AppGamma}
%%%%%%%%%%%%%%%%%%%%%%%%%%%%%%%%%%%%%

{
As stated in the main text, our criterion for identifying the phase of QCD matter in NS cores is based on analyzing the behavior of the polytropic index $\gamma \equiv \mathrm{d}(\ln p)/\mathrm{d}(\ln \epsilon)$, \emph{i.e.},~the slope of the EoS in \fig{1}  and Extended Data  Figs.~1, panel a; 2; and 3.  Here, we comment on the physics behind this statement, and explain our choice to identify the presence of quark matter using as the quantitative criterion $\gamma<1.75$ continuously up to asymptotic densities. 

Matter that exhibits exact conformal symmetry, \emph{i.e.},~does not possess intrinsic mass scales, is characterized by $\gamma=1$ independent of the strength of the coupling. This is so because in the absence of any dimensionful parameters, the energy density and pressure must to be proportional to each other, leading to $\gamma=1$. This symmetry can also be shown to lead to a speed of sound squared $c_s^2 = 1/3$ for the system.

In low- and moderate-density QCD matter, it is well known that the ground state does not exhibit the approximate chiral symmetry of the underlying Lagrangian (see \emph{e.g.} ref.~\cite{Holt:2009ty}\ for details). This spontaneous breaking of the symmetry leads to the emergence of the fundamental scales of nuclear matter, such as hadron masses, and scale-dependent interactions.
These mass scales lead to a highly non-conformal behavior for the EoS, which is reflected in the polytropic index taking large values, typically in excess of 2, in viable models of high-density nuclear matter. 

Collections of $\gamma$ values that different nuclear physics models predict are available through the related adiabatic index $\Gamma=\frac{\epsilon+p}{\epsilon}\,\gamma$, tabulated \emph{e.g.}~in Table III of ref.~\cite{Read:2008iy}\ and plotted in figure 5.9 of ref.~\cite{Haensel:2007yy}. A closer inspection of the wide class of EoSs we have gathered from refs.~\cite{Lattimer:2000nx,Gandolfi:2011xu,Fortin:2016hny} shows that while there are a number of EoSs for which the polytropic index reaches values of order 1.5 or below, all of these are in conflict with the recent LIGO/Virgo tidal deformability bound---a constraint that in particular rules out typical hyperonic EoSs. For the viable hadronic EoSs we have analyzed, $\gamma$ stays around or above 1.75 in all cases except for MPA1 for which the parameter can drop to approx.~1.6 at very high densities. This EoS, however, exhibits a speed of sound squared extremely close to unity exactly when $\gamma$ falls below 2. In addition to casting doubt on its reliability, this fact highlights the lack of overlap between its high-density behavior and that of our family of interpolated EoSs.

In high-density quark matter, on the other hand, the underlying approximate chiral symmetry of QCD is restored, and, as a result, the system exhibits approximate conformal symmetry. Minor violations of conformal behavior arise from the masses of the up, down, and strange quarks, which are, however, very small compared to the nucleon masses. Moreover, the interactions between quarks and gluons also lead to a mild breaking of conformal symmetry in the quark-matter phase, manifesting as a logarithmic dependence of the strong coupling constant on the baryon chemical potential. To a good accuracy, however, quark matter always behaves as an ultrarelativistic gas of interacting quarks and gluons, which becomes even more pronounced in the controlled, perturbative high-density region of the QCD phase diagram, where the polytropic index $\gamma$ quickly approaches unity.

In the most nontrivial density range near the deconfinement transition, QCD matter evolves from the highly nonconformal hadronic behavior to one characteristic of quark matter. This transition may take place either as a discontinuous jump in energy density, in which case the value $\gamma=1.75$ may never be reached, or in a smooth crossover manner, whereby a crisp phase identification may not always be feasible (then there may even exist an overlap region where the system can be described both in terms of nuclear and quark degrees of freedom). In either case, our results indicate that the cores of typical pulsars and maximal-mass NSs very likely do not reside here, but rather safely belong to the nuclear and quark matter regimes, respectively. This is reflected in the fact that our qualitative conclusions are not sensitive to the exact choice of the critical polytropic index as long as it resides between the hadronic and quark matter regimes. Indeed, only our detailed numerical conclusions would be somewhat modified, should we vary the number $\gamma=1.75$ in a moderate way.

}

%%%%%%%%%%%%%%%%%%%%%%%%%%%%%%%%%%%%%
\subsection{Analysis of EoS smoothness}
\label{AppSmoothing}
%%%%%%%%%%%%%%%%%%%%%%%%%%%%%%%%%%%%%

In addition to {reaching} large speeds of sound, one way in which some of the EoSs generated by the speed-of-sound interpolation method can be classified as extreme is that the piecewise nature of the interpolation functions allows for very quick changes in the material properties of the matter in arbitrarily small density windows. While such versatility is in principle a desirable feature of the interpolator, these structures are not very likely to appear in Nature, and in addition bring unnecessary complications to the polytropic-index analysis performed in the main text. To quantify the level of local structure in our EoSs, we classify them according to the smallest (logarithmic) energy-density interval where structures appear. In practice, this is implemented by demanding that the energy densities at two successive inflection points $\epsilon_i$ and $\epsilon_{i+1}$, where the speed of sound changes its behavior, satisfy $(\epsilon_{i+1}- \epsilon_i)/\epsilon_i > \DeltaLogEps$ with a given constant $\DeltaLogEps > 0$.  Note that imposing this constraint does not exclude discontinuous first-order phase transitions or rapid crossovers.  

{As demonstrated in Extended Data \fig{2}, 
we find that placing minor smoothness limits} ($\DeltaLogEps \lesssim 1$) affects the allowed EoS region mainly around the matching points, where the EoS is best known, but does not have a significant effect at intermediate densities.  However, somewhat larger values ($\DeltaLogEps \gtrsim 1 $) begin to significantly constrain the allowed region at all densities. This shows that the EoSs that make up the boundaries of the {our EoS band} must exhibit both very large speeds of sounds as well as rapid changes in material properties.

For our analysis of the central values of $\gamma$ in stars of different masses, we have used $\DeltaLogEps >0.5$, which 
has a minor effect on the global characteristics of the EoS family. { In particular, this cut has a minor effect on the mass-radius relation for stars above $1.4 M_\odot$, shifting the extremes of the allowed radius for a given mass by about $0.3$~km at most.} Moreover, we note that for completeness we have also allowed the first inflection point to approach $n_\text{CET}$ without limit, finding that all the results presented in the main text remain unchanged.

%%%%%%%%%%%%%%%%%%%%%%%%%%%%%%%%%%%%%
\subsection{Comparison with recent mass and radius constraints}
\label{AppMR}
%%%%%%%%%%%%%%%%%%%%%%%%%%%%%%%%%%%%%

{

Our ensemble of EoSs can also be transformed to the $M$-$R$ plane in order to compare its behaviour to recent radius observations of NSs. In Extended Data \fig{5}, 
we overlay a few representative X-ray MR-measurements on top of our family of MR-curves that are obtained from the EoS ensemble of 
\fig{1}.
We show examples of measurements obtained with three different methods: 
direct atmosphere-model fits to the time-evolving X-ray burst spectra (corresponding to the low-mass X-ray binary [LMXB] system 4U 1702$-$429 [yellow curve])\cite{Nattila:2017wtj},  cooling-tail method fits to X-ray burst observations (LMXBs 4U 1724$-$307 [light brown] and SAX J1810.8$-$2609 [cyan])\cite{Nattila:2015jra}, and quiescent LMXB spectra fits to sources with reliable distance measurements (NGC 6304 [dark brown], NGC 6397 [green], M13 [purple], M28 [orange], M30 [black], $\omega$ Cen [magenta], 47 Tuc X5 [blue], and 47 Tuc X7 [red])\cite{Steiner:2010fz, Guillot:2013wu,Ozel:2015fia,Bogdanov:2016nle,Steiner:2017vmg,Shaw:2018wxh}.
For the quiescent LMXB measurements we use public data from refs.~\cite{Steiner:2017vmg, Shaw:2018wxh}.  
We assume, for simplicity, no hot spots.  
The corresponding atmospheres are assumed to be composed of either hydrogen (dotted lines) or helium (dashed lines), depending on the source.

In general this kind of qualitative comparison remains largely inconclusive and warrants a further quantitative treatment taking into account the full interplay between different measurements and their uncertainties.
That being said, we note that the measurements are consistent with the lower $c_s^2$ values as the measured radii are typically around $R \approx 12~\mathrm{km}$.
This is especially true for the most precise MR measurement concerning the NS in 4U 1702$-$429, which is fully compatible with the subconformal-EoS region, where $c_s^2 < 1/3$.
We emphasize that this particular selection of measurements presented in 
%Fig.~\ref{EDfig5} 
Extended Data \fig{5} 
is by no means meant to be exhaustive.
A more detailed self-consistent Bayesian treatment of the problem with all the available measurements present in the literature is left for future work.

}

\end{methods}

\begin{addendum}
    \item[Data Availability] 
    Source data are available for this paper. For three tabulated sample EoSs and the boundary regions for the $p(\epsilon)$ (Fig.~1) and MR regions (Extended Data \fig{5}), please consult the Supplemental files and Source Data files. 
    The authors will in addition be happy to provide more EoS tables upon request.

\end{addendum}

\begin{addendum}
    \item[Code availability] The code used to construct the interpolated EoSs is available from the authors upon request.  
\end{addendum}

\begin{addendum}
    \item[Author contributions] The authors conducted the research together and participated in the preparation and revision of the manuscript. The speed-of-sound interpolation was implemented by T.G.~and the spectral interpolation by E.A.,\ both based on an earlier code written by A.K.  
\end{addendum}

\begin{addendum}
    \item[Competing interests] The authors declare no competing interests.  
\end{addendum}

%%%%%%%%%%%%%%%%%%%%%%%%%%%%%%%%%%%%%%
\begin{figure}
\caption{
\label{EDfig1}
\textbf{A comparison of the three interpolation methods considered.} Ranges of allowed equations of state (\textbf{a}) and the corresponding neutron star mass-radius relations (\textbf{b}) are displayed as obtained with the speed-of-sound (green region), piecewise-polytropic (black dotted lines), and spectral (orange dashed lines) interpolation methods. In panel \textbf{b}, we show also EoSs that fail to reproduce 2$M_\odot$ stars (cyan color) and those that lead to tidal deformabilities outside the range determined by the gravitational-wave event GW170817\cite{Abbott:2018exr} (purple).}{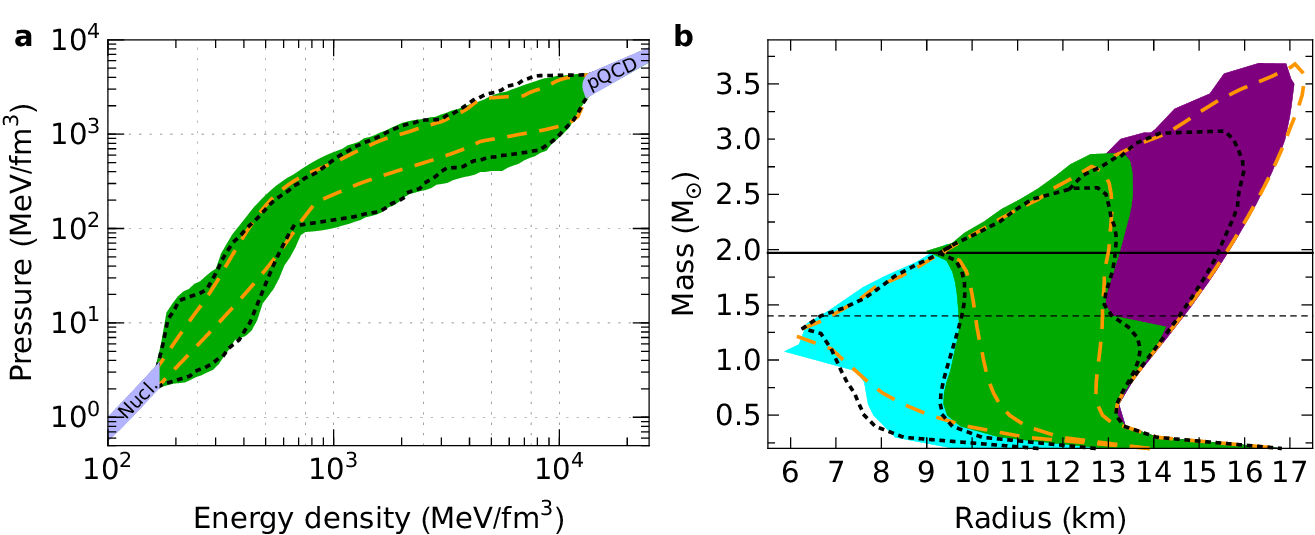}{1\textwidth}
\end{figure}
%%%%%%%%%%%%%%%%%%%%%%%%%%%%%%%%%%%%%%%%%%%%%%%%%%%%%%%

%%%%%%%%%%%%%%%%%%%%%%%%%%
\begin{figure}
\caption{
\label{EDfig2}
\textbf{Limiting the amount of fine structure in the equations of state.} We display EoS bands subjected to the additional constraint of setting a lower limit for the logarithmic segment size in energy density, $\DeltaLogEps$. In our analysis of the polytropic-index distributions, we have applied the lower limit $\DeltaLogEps > 0.5$, which affects the EoS band mainly near the low- and high-density limits. 
}
{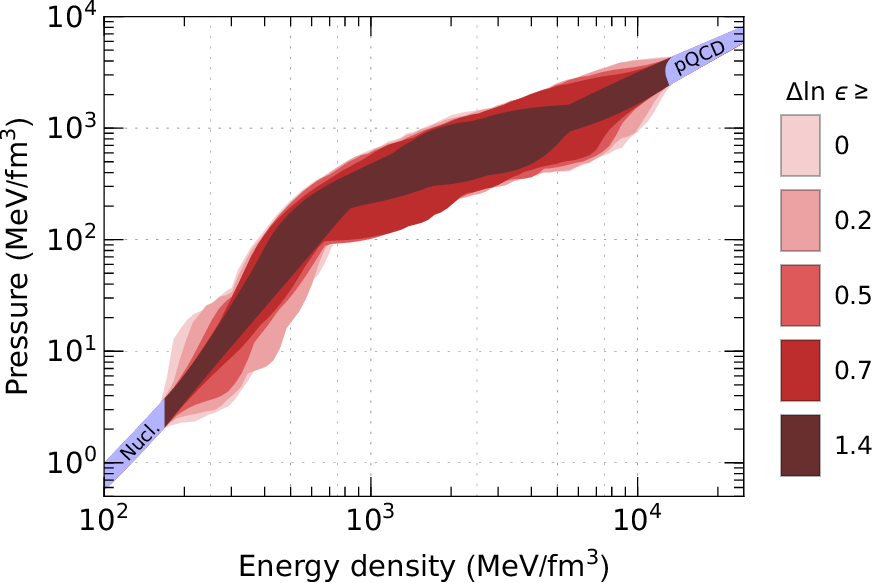}{0.7\textwidth}
\end{figure}
%%%%%%%%%%%%%%%%%%%%%%%%

%%%%%%%%%%%%%%%%%%%%%%%%
\begin{figure}
\caption{
\label{EDfig3}
\textbf{The central densities of neutron stars of different masses.} The central densities and pressures of different stars are displayed on top of a background showing the ranges of possible NS matter EoSs. Maximally massive NSs  (red dots), 2$M_{\odot}$ NSs (orange squares), and 1.44$M_{\odot}$ NSs (blue diamonds) are shown on top of each other. The color coding of the background refers to the maximal value that the speed of sound squared $c_{s}^{2}$ reaches at any density.}{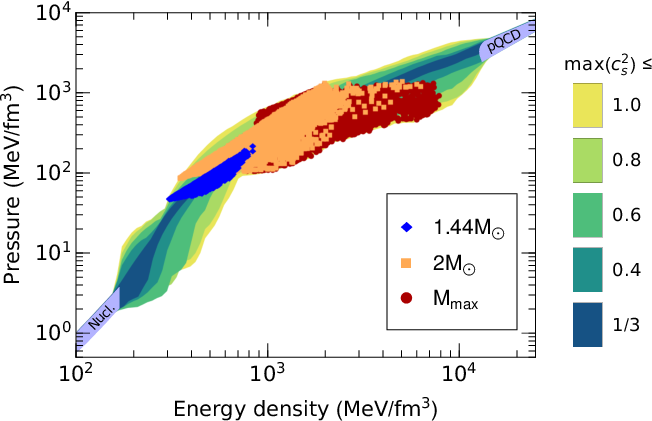}{0.7\textwidth}
\end{figure}
%%%%%%%%%%%%%%%%%%%%%%%%%%%%%%%%%%%%%%%%%%%%%%%%%%%%%%%

%%%%%%%%%%%%%%%%%%%%%%%%%%%%%%%%%%%%%
\begin{figure}
\caption{
\label{EDfig4}
\textbf{The size of the quark core in  two-solar-mass neutron stars.} The extent of the quark-matter cores in $2 M_\odot$ neutron stars as a function of the maximal mass $M_{max}$ for each equation of state. The color coding of the points refers to the maximal value that the speed of sound squared $c_{s}^{2}$ reaches at any density for the corresponding equations of state.}{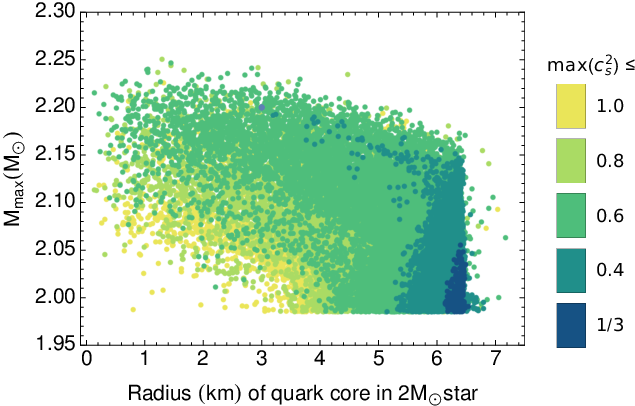}{0.7\textwidth}
\end{figure}
%%%%%%%%%%%%%%%%%%%%%%%%%%%%%%%%%%%%%

\begin{figure}
\caption{
\label{EDfig5}
{\textbf{Comparison of mass-radius predictions to recent observations.} 
We compare the MR-curves obtained with our interpolated equations of state to several recent simultaneous MR-measurements. X-ray burst constraints are indicated by solid lines and quiescent low-mass X-ray binary system measurements by dotted (hydrogen atmosphere) or dashed (helium atmosphere) lines.
We show examples of measurements obtained with three different methods: 
direct atmosphere-model fits to the time-evolving X-ray burst spectra (corresponding to the low-mass X-ray binary [LMXB] system 4U 1702$-$429 [yellow curve])\cite{Nattila:2017wtj},  cooling-tail method fits to X-ray burst observations (LMXBs 4U 1724$-$307 [light brown] and SAX J1810.8$-$2609 [cyan])\cite{Nattila:2015jra}, and quiescent LMXB spectra fits to sources with reliable distance measurements (NGC 6304 [dark brown], NGC 6397 [green], M13 [purple], M28 [orange], M30 [black], $\omega$ Cen [magenta], 47 Tuc X5 [blue], and 47 Tuc X7 [red])\cite{Steiner:2010fz, Guillot:2013wu,Ozel:2015fia,Bogdanov:2016nle,Steiner:2017vmg,Shaw:2018wxh}.
For the quiescent LMXB measurements we use public data from refs.~\cite{Steiner:2017vmg, Shaw:2018wxh}. 
We assume, for simplicity, no hot spots.  
The corresponding atmospheres are assumed to be composed of either hydrogen (dotted lines) or helium (dashed lines), depending on the source. The color coding of the points refers to the maximal value that the speed of sound squared $c_{s}^{2}$ reaches at any density for the corresponding equations of state.}
}{extended_data_fig5}{0.7\textwidth}
\end{figure}

%%%%%%%%%%%%%%%%%%%%%%%%%%%%%%%%%%%%%%%%%%%%%%%%%%%%%

\clearpage
%\bibliography{tmp-references.bib}

\end{document}